\documentclass[aps,prb,twocolumn,superscriptaddress,showpacs]{revtex4}

\usepackage{graphicx}
\usepackage{color}

\newcommand{\lcmo}{La$_{0.5}$Ca$_{0.5}$MnO$_{3}$}
\newcommand{\lpcmo}{La$_{0.25}$Pr$_{0.375}$Ca$_{0.375}$MnO$_{3}$}
\newcommand{\icm}{cm$^{-1}$}

\begin{document}

\title{Raman scattering studies of temperature- and field-induced melting of charge order in
(La,Pr,Ca)MnO$_{3}$}

\author{M. Kim}
\author{H. Barath}
\author{S.L. Cooper}
\author{P. Abbamonte}
\author{E. Fradkin}
\affiliation{Department of Physics and Frederick Seitz Materials
Research Laboratory, University of Illinois, Urbana, Illinois 61801}
\author{M. R{\"u}bhausen}
\affiliation{Institut f{\"u}r Angewandte Physik, Universit{\"a}t
Hamburg, Jungiusstra{\ss}e 11, D-20355 Hamburg, Germany}
\author{C. L. Zhang}
\author{S-W. Cheong}
\affiliation{Rutgers Center for Emergent Materials and Department of
Physics and Astronomy, Rutgers University, Piscataway, New Jersey
 08854, USA}

\date{\today}

\begin{abstract}
We present Raman scattering studies of the structural and magnetic
phases that accompany temperature- and field-dependent melting of
charge- and orbital-order (COO) in La$_{0.5}$Ca$_{0.5}$MnO$_{3}$ and
La$_{0.25}$Pr$_{0.375}$Ca$_{0.375}$MnO$_{3}$.  Our results show that
thermal and field-induced COO melting in
La$_{0.5}$Ca$_{0.5}$MnO$_{3}$ exhibits three stages in a
heterogeneous melting process:  at low temperatures and fields, we
observe a long-range, strongly Jahn-Teller (JT) distorted/COO phase;
at intermediate temperatures and/or fields, we find a coexistence
regime comprising both strongly JT distorted/COO and weakly JT
distorted/ferromagnetic metal (FMM) phases; and at high temperatures
and/or high fields, we observe weakly JT distorted homogeneous
paramagnetic (PM) or ferromagnetic (FM) phase. In the high
field/high temperature regime of La$_{0.5}$Ca$_{0.5}$MnO$_{3}$ and
La$_{0.25}$Pr$_{0.375}$Ca$_{0.375}$MnO$_{3}$, we identify a clear
structural change to a weakly JT distorted phase that is associated
with either a \textit{Imma} or \textit{Pnma} structure.
 We are able to provide a complete structural phase diagram of
La$_{0.5}$Ca$_{0.5}$MnO$_{3}$ for the temperature and field ranges
6${\leq}$$T$${\leq}$170 K and 0${\leq}$$H$${\leq}$9 T.
Significantly, we provide evidence that the field-induced melting
transition of La$_{0.5}$Ca$_{0.5}$MnO$_{3}$ is first-order, and
resembles a crystallization transition of an ``electronic solid.''
We also investigate thermal and field-induced melting in
La$_{0.25}$Pr$_{0.375}$Ca$_{0.375}$MnO$_{3}$ to elucidate the role
of disorder in melting of COO.  We find that while thermal melting
of COO in La$_{0.25}$Pr$_{0.375}$Ca$_{0.375}$MnO$_{3}$ is quite
similar to that in La$_{0.5}$Ca$_{0.5}$MnO$_{3}$, field-induced
melting of COO in the two systems is quite different in several
respects:  the field-induced transition from the COO phase to the
weakly JT-distorted/FM phase in
La$_{0.25}$Pr$_{0.375}$Ca$_{0.375}$MnO$_{3}$ is very abrupt, and
occurs at significantly lower fields ($H$$\sim$2 T at $T$$\sim$0 K)
than in La$_{0.5}$Ca$_{0.5}$MnO$_{3}$ ($H$$\sim$30 T at $T$=0 K);
the intermediate coexistence regime is much narrower in
La$_{0.25}$Pr$_{0.375}$Ca$_{0.375}$MnO$_{3}$ than in
La$_{0.5}$Ca$_{0.5}$MnO$_{3}$; and the critical field $H_{c}$
increases with increasing temperature in
La$_{0.25}$Pr$_{0.375}$Ca$_{0.375}$MnO$_{3}$, in contrast to the
decrease in $H_{c}$ observed with increasing temperature in
La$_{0.5}$Ca$_{0.5}$MnO$_{3}$.  To explain these differences, we
propose that field-induced melting of COO in
La$_{0.25}$Pr$_{0.375}$Ca$_{0.375}$MnO$_{3}$ is best described as
the field-induced percolation of FM domains, and we suggest that
Griffiths phase physics may be an appropriate theoretical model for
describing the unusual temperature- and field- dependent transitions
observed in La$_{0.25}$Pr$_{0.375}$Ca$_{0.375}$MnO$_{3}$.
\end{abstract}

\pacs{73.43.Nq, 75.47.Lx, 78.30.-j}

\maketitle

\section{Introduction}
Strong phase competition in the manganese perovskites
$A_{1-x}B_{x}$MnO$_{3}$ ($A$=trivalent rare earth, $B$=divalent
alkaline earth) is known to spawn many of the interesting properties
of these materials, such as their diverse phase behaviors as
functions of temperature, rare earth ionic radius, doping, magnetic
field, and
pressure,~\cite{schiffer95,kuwahara95,tomioka95,asamitsu95,kawano96,tomioka96,hwang95,fontcuberta96,radaelli97,rao97,laukhin97}
as well as the exotic phenomena these materials exhibit, including
electronic phase separation,~\cite{fath99,moritomo99,uehara99,rao03}
colossal
magnetoresistance,~\cite{jin94,tokura96,ramirez99,salamon01} and
charge/orbital ordering (COO).~\cite{wollan55,jirak85,rao00}  For
example, La$_{0.5}$Ca$_{0.5}$MnO$_{3}$ exhibits a transition from a
high temperature paramagnetic insulating (PI) phase to a
ferromagnetic metal (FMM) phase at $T_{C}$=225 K, and then to a
CE-type antiferromagnetic insulating (AFI) phase below $T_{N}$=155
K, in which the $e_g$ orbitals on the Mn$^{3+}$ sites alternate
between $d_{3x^2-r^2}$ and $d_{3z^2-r^2}$ orbital configurations
lying in the $a$-$c$ plane (in the $Pnma$
representation),~\footnote{In the $Pbnm$ representation,
$d_{3x^2-r^2}$ and $d_{3y^2-r^2}$ orbitals alternate lying in the
$a$-$b$ plane.} and the equal number of Mn$^{3+}$ and Mn$^{4+}$ ions
form a checkerboard pattern on the
lattice.~\cite{schiffer95,wollan55,radaelli95,radaelli9755,goodenough55}
 La$_{0.5}$Ca$_{0.5}$MnO$_{3}$ also exhibits significant structural
changes associated with its electronic/magnetic phase transitions.
 At room temperature, the structure of La$_{0.5}$Ca$_{0.5}$MnO$_{3}$
is described by the space group
\textit{Pnma}.~\cite{radaelli9755,huang00}  However, below $T_{C}$
there is evidence for the development of nanometer-scale phase
regions with a superlattice structure having the space group
\textit{P$2_1$/m}, in which the unit cell is doubled along the $a$
axis.~\cite{chen96}  These \textit{P$2_1$/m} regions become more
prominent with decreasing temperature below $T_{C}$, and eventually
occupy the entire structure of La$_{0.5}$Ca$_{0.5}$MnO$_{3}$ below
$T_{N}$.~\cite{mori98}
 There are also large changes in the lattice parameters between
$T_{C}$ and $T_{N}$ in La$_{0.5}$Ca$_{0.5}$MnO$_{3}$:  as the
temperature decreases from $T_{C}$ to $T_{N}$, the $a$ axis
increases by ${\Delta}a$$\sim$0.022 {\AA}, the $c$ axis increases by
${\Delta}c$$\sim$0.041 {\AA}, and the $b$ axis decreases by
${\Delta}b$$\sim$$-$0.122
{\AA}.~\cite{wollan55,radaelli95,radaelli9755}  This complex
transition is associated with CE-type AFM charge/orbital
order.~\cite{wollan55,radaelli95,radaelli9755}  The zig-zag pattern
of the COO configuration in La$_{0.5}$Ca$_{0.5}$MnO$_{3}$ (see Fig.
1(b))~\cite{wollan55,jirak85,rao00,radaelli95,radaelli9755} is
associated with cooperative JT distortions of the Mn$^{3+}$O$_{6}$
octahedra that lower the energy of those orbital states having lobes
oriented along the long Mn-O octahedral bond directions.
 Accordingly, the polarization of orbitals in the \textit{a-c} plane
accounts for the decrease of the \textit{b} axis---and the increase
of the \textit{a} and \textit{c} axes---below $T_{N}$.  The COO
configuration in La$_{0.5}$Ca$_{0.5}$MnO$_{3}$ leads to an
inequivalency of adjacent Mn sites below $T_{N}$, resulting in a
doubling of the unit
cell,~\cite{radaelli95,radaelli9755,chen96,abrashev01,iliev01} and
favors CE-type AFM ordering, in which the spins \textit{within} the
zigzag chains are ferromagnetically ordered, while spins between
adjacent chains are antiferromagnetically ordered. This magnetic
ordering pattern is suggested by the semicovalent exchange coupling
picture of Goodenough:~\cite{goodenough55}  the singly
semicovalent-bonded Mn$^{3+}$-O-Mn$^{4+}$ structure along the chains
favors FM coupling, while the doubly semicovalent-bonded
Mn$^{3+}$-O-Mn$^{3+}$/Mn$^{4+}$-O-Mn$^{4+}$ structure between the
adjacent chains favors AFM coupling.~\cite{goodenough55}

A topic that continues to be of great interest in studies of COO in
the manganites concerns the manner in which the COO state melts into
disordered phases, either via classical thermal melting, or via
quantum mechanical ($T$$\sim$0) melting.  Temperature-dependent
melting of COO in the manganites has been well-studied in a variety
of measurements.  For example, magnetic and transport measurements
show that COO in La$_{0.5}$Ca$_{0.5}$MnO$_{3}$ melts into a FM metal
phase above $T_N$, and that this transition is
hysteretic---exhibiting an FM to AFM transition at $T_N$=135 K on
cooling, and at $T_N$=180 K on warming---indicating that the
thermally driven transition between AFM insulating and FM metal
phases is first-order.~\cite{schiffer95}  Electron and x-ray
diffraction measurements of La$_{0.5}$Ca$_{0.5}$MnO$_3$ further show
that the AFM to FM transition coincides with a transition from a
commensurate (CM) to incommensurate (IC) charge-ordered phase, in
which the COO superlattice peaks exhibit a decreased
commensurability with the lattice and a suppression in intensity
with increasing temperature up to
$T_{CO}$=240K.~\cite{radaelli95,radaelli9755,chen96}  These results
suggest that IC charge ordering coexists with the ferromagnetic
metal phase in the temperature regime between $T_N$ and $T_{CO}$.
 Electron microscopy studies also show that two-phase coexistence in
La$_{0.5}$Ca$_{0.5}$MnO$_3$, La$_{5/8-y}$Pr$_y$Ca$_{3/8}$MnO$_3$
($y$=0.375, and 0.4), and La$_{0.33}$Ca$_{0.67}$MnO$_3$ originates
from an inhomogeneous spatial mixture of FM and CO domains, whose
sizes vary with temperature.~\cite{uehara99,mori98,tao04}  It was
proposed that IC charge ordering results from thermal disordering of
the \textit{e$_g$} orbitals in the CO regions, which finally
collapse completely at $T_{CO}$.~\cite{mori98}  The same mechanism
for thermal melting was also observed in electron diffraction and
electron microscopy measurements of La$_{0.45}$Ca$_{0.55}$MnO$_3$
and Pr$_{0.5}$Ca$_{0.5}$MnO$_3$.~\cite{chen99}

In contrast to thermal melting of COO in the manganites, quantum
mechanical ($T$=0) melting of COO in the manganites has been much
less well studied, particularly using microscopic diffraction or
spectroscopic probes.  In quantum mechanical melting, the disruption
of COO is induced near $T$=0 by tuning a control parameter such as
doping, pressure, and magnetic field; consequently, quantum, rather
than thermal, fluctuations may play a significant role in quantum
melting of
COO.~\cite{kuwahara95,tomioka95,tomioka96,okimoto99,naler01}  Most
studies of quantum mechanical melting behavior have relied upon
chemical substitution at the divalent cation site, but this method
of phase tuning creates a disorder potential that is expected to
have significant effects near quantum phase
transitions.~\cite{schiffer95,tomioka96,moritomo99,hwang95,naler01,roy98}
 While more limited in number, studies of magnetic-field-tuned
melting of charge order in the manganites have been revelatory:
 Transport and magnetic studies of magnetic-field-tuned COO melting
in Nd$_{0.5}$Sr$_{0.5}$MnO$_3$, Pr$_{0.5}$Sr$_{0.5}$MnO$_3$, and
Pr$_{1-x}$Ca$_x$MnO$_3$ ($x$=0.3, 0.35, 0.4, and 0.5) showed that an
applied magnetic field disrupts the AFM order associated with COO,
and thereby disrupts the COO state, leading to a sharp drop in
resistivity and a transition to a FM metal
phase.~\cite{kuwahara95,tomioka95,tomioka96,okimoto99}  The
hysteresis observed in these measurements indicates that the
field-tuned AFM-to-FM transition is also a first order
transition.~\cite{kuwahara95,tomioka95,tomioka96,okimoto99}  Optical
measurements of Pr$_{0.6}$Ca$_{0.4}$MnO$_3$ and
Nd$_{0.5}$Sr$_{0.5}$MnO$_3$ have also investigated the effects of
thermal and field-induced melting of charge/orbital order on the
optical spectral weight over a wide energy
range.~\cite{jung00,okimoto99}  However, while these bulk
measurements have provided important information regarding the
effects of thermal and field-induced melting on the physical
properties of the manganites, they convey little information
regarding the structural or other microscopic changes that accompany
quantum melting of COO in the manganites.

Structural studies of COO field-induced melting in the manganites
have been very limited.  For example, field-dependent x-ray
diffraction measurements of La$_{0.5}$Ca$_{0.5}$MnO$_3$, performed
in the field range $H$=0 to 10 T, have been performed only at
$T$=115 K, showing that the variance of the Mn-O-Mn bond
length---and hence the size of the JT distortion---decreases with
increasing field at this temperature.~\cite{tyson04}  However, there
have been, as yet, no extensive structural studies of field-induced
COO melting in the manganites over a more complete range of the
$H$-$T$ phase diagram.

The dearth of microscopic measurements of field- and pressure-tuned
quantum phase transitions associated with COO in the manganites has
left unanswered a variety of important questions, including:  What
microscopic changes accompany quantum mechanical melting of COO in
the manganites as functions of pressure and magnetic field?  Is
there evidence in quantum phase transitions for novel phase
behavior, such as electronic liquid crystal~\cite{kivelson98} or
spin-glass phases? What magnetic and structural phases result when
COO melts by tuning magnetic field or pressure?  What is the role of
disorder in the field- or pressure-tuned quantum phase transitions?

Field- and pressure-tuned inelastic light (Raman) scattering
measurements provide a powerful method for studying structural and
other microscopic details associated with quantum phase transitions
in complex oxides.~\cite{snow02,snow03,karpus04}  For example, Raman
scattering can provide energy, lifetime, and symmetry information
about the important spin, charge, lattice, and orbital
degrees-of-freedom in the manganites, and can convey important
information regarding structural changes that accompany temperature-
and field-dependent phase changes.  Additionally, Raman measurements
can be readily made under the `extreme' conditions of low
temperature, high magnetic field, and/or high pressure needed for
studying quantum phase transitions in correlated
materials.~\cite{snow02,snow03,karpus04}

In this paper, we present temperature- and field-dependent Raman
measurements of La$_{0.5}$Ca$_{0.5}$MnO$_3$ and
La$_{0.25}$Pr$_{0.375}$Ca$_{0.375}$MnO$_{3}$, which enable us to map
out the structural and COO phases of these materials in the
temperature range 2--300 K and in the field range 0--9 T.  Among the
interesting results of these measurements:  (i) We observe strong
evidence that both field- and temperature-dependent melting of COO
in La$_{0.5}$Ca$_{0.5}$MnO$_3$ occurs in such a way as to preserve
large coherent domains of COO throughout the melting process,
indicating that field- and temperature-dependent melting in this
material occurs primarily at the surfaces of COO regions.  These
results provide evidence that the field-induced melting transition
of La$_{0.5}$Ca$_{0.5}$MnO$_3$ is first-order, and resembles a
crystallization transition of an ``electronic solid.''  (ii) The
similar effects of field- and temperature-dependent melting on the
structure of La$_{0.5}$Ca$_{0.5}$MnO$_3$ further suggests that
magnetic fields and thermal disorder disrupt charge-ordering through
a similar mechanism, i.e., by locally disordering AFM alignment.
 (iii) We observe the appearance of new modes in the high-field
regimes of both La$_{0.5}$Ca$_{0.5}$MnO$_3$ and
La$_{0.25}$Pr$_{0.375}$Ca$_{0.375}$MnO$_{3}$, which indicate that
the melted high-field FM phase of these materials is associated with
a structural phase having weakly JT distorted MnO$_6$ octahedra.
 (iv) Finally, we find that substitution of Pr into
La$_{1-x}$Ca$_{x}$MnO$_3$---which allows us to study the effects of
disorder on thermal and field-induced melting of the COO
lattice---has dramatically different effects on thermal and
field-induced COO melting. Specifically, we find that while thermal
melting of COO into La$_{1-x}$Ca$_{x}$MnO$_3$ is not dramatically
affected by Pr-substitution, field-induced melting in
La$_{0.25}$Pr$_{0.375}$Ca$_{0.375}$MnO$_{3}$ is substantially more
abrupt than that in La$_{0.5}$Ca$_{0.5}$MnO$_3$.  To explain this,
we propose a Griffiths phase picture of COO melting in
La$_{0.25}$Pr$_{0.375}$Ca$_{0.375}$MnO$_{3}$ in which COO melting
occurs via field-induced percolation of FM domains.

\section{Experimental procedure}
\subsection{Sample details and preparation}
Raman scattering measurements were performed on the
as-grown surfaces of high-purity polycrystalline pellets of
La$_{0.5}$Ca$_{0.5}$MnO$_3$ ($T_{CO}$=240 K;
Ref.~\onlinecite{chen96}) and
La$_{0.25}$Pr$_{0.375}$Ca$_{0.375}$MnO$_{3}$ ($T_{CO}$=210 K;
Ref.~\onlinecite{uehara99}).  The pellets of both materials
contained typical crystallite sizes between
3${\times}$3${\times}$3${\mu}$m$^3$ and
5${\times}$5${\times}$5${\mu}$m$^3$.

In the paramagnetic phase, La$_{0.5}$Ca$_{0.5}$MnO$_3$ has an
orthorhombic structure associated with the \textit{Pnma} space
group, which is isostructural to
LaMnO$_3$.~\cite{radaelli97,abrashev01,iliev01}  The \textit{Pnma}
structure is that of the simple cubic perovskite structure (
\textit{$Pm\overline{3}m$} group symmetry), but with the MnO$_6$
octahedra Jahn-Teller distorted and rotated along the [010] and
[101] cubic axes.  These changes give rise to the following values
for the lattice parameters in La$_{0.5}$Ca$_{0.5}$MnO$_3$:
 $a{\approx}b{\approx}{\sqrt{2}}a_c$ and $c{\approx}2a_c$, where
$a_c$=the cubic perovskite lattice
parameter.~\cite{abrashev01,iliev01,iliev98}  Below $T_N$, x-ray and
electron diffraction measurements indicate that the symmetry of
La$_{0.5}$Ca$_{0.5}$MnO$_3$ is lowered in the COO phase to a
$P2_1/m$ structure, which involves a doubling of the \textit{Pnma}
unit cell to a structure with lattice parameters $a{\approx
}2\sqrt{2}a_c$, $b{\approx}\sqrt{2}a_c$ and
$c{\approx}2a_c$.~\cite{radaelli9755,chen96}

\subsection{Raman measurements}
The Raman scattering measurements were performed using the 647.1 nm
excitation line from a Kr$^+$ laser.  The incident laser power was
limited to 10 mW, and was focused to a ${\sim}$50${\mu}$m-diameter
spot to minimize laser heating of the sample.  The scattered light
from the samples was collected in a backscattering geometry,
dispersed through a triple stage spectrometer, and then detected
with a liquid-nitrogen-cooled CCD detector.  The incident light
polarization was selected with a polarization rotator, and the
scattered light polarization was analyzed with a linear polarizer,
providing symmetry information about the excitations studied.  The
samples were inserted into a continuous He-flow cryostat, which was
itself mounted in the bore of a superconducting magnet, allowing
measurements in both the temperature range 4--350 K and the
magnetic-field range 0--9 T.

\subsection{Field measurements}
Magnetic field measurements were performed in the Faraday
($\overrightarrow{q}$$\parallel$\textbf{H}) geometry.  To avoid
Faraday rotation effects in applied magnetic fields, the incident
light was circularly-polarized in the (\textbf{E}$_i$,
\textbf{E}$_s$)=(\textbf{L}, \textbf{R}) geometry,~\footnote{In this
cross configuration of the incident and scattered light
polarizations, i.e., (\textbf{E}$_i$, \textbf{E}$_s$)=(\textbf{L},
\textbf{R}), the $A_g$, $B_{1g}$, $B_{2g}$, and $B_{3g}$ symmetries
are allowed because the diagonal elements of the Raman tensor,
which represent the $A_g$ symmetry, are not equal, i.e.,
${\alpha}_{xx}{\neq}{\alpha}_{yy}{\neq}{\alpha}_{zz}$.  While all
the symmetries are allowed in the (\textbf{E}$_i$, \textbf{E}$_s$)
=(\textbf{L}, \textbf{R}) configuration, the previous observations
indicate that the $B_{1g}$ and $B_{3g}$ symmetry modes in $Pnma$
($B_{2g}$ and $B_{3g}$ in $Pbnm$) are very weak,~\cite{abrashev01}
so that most of the observed modes have either $A_g$ or $B_{2g}$
($B_{1g}$ in $Pbnm$) symmetry.} where \textbf{E}$_i$ and
\textbf{E}$_s$ are the incident and scattered electric field
polarizations, respectively, and \textbf{L(R)} stands for
left(right) circular polarization.  This geometry allowed us to
investigate modes with symmetries $A_g+B_{1g}+B_{2g}+B_{3g}$.  The
incident light was converted from linearly to circularly polarized
light using a Berek compensator optimized for the 647.1 nm line, and
the scattered light was converted back to linearly polarized light
with a quarter-wave plate.  In this paper, when listing the
excitation symmetries observed, rather than referring to the
$P2_1/m$ space group of the actual structure, we use the irreducible
representations of the orthorhombic $Pmma$ space group of the
simplified structure in which the octahedral tilts (rotations) are
not present, but in which the symmetry is lowered due to
charge/orbital ordering. This simplification is
justified~\cite{abrashev01} by the observation that the most intense
Raman lines observed in La$_{0.5}$Ca$_{0.5}$MnO$_3$ are associated
with those observed in the COO layered manganites.~\cite{yamamoto00}

\begin{figure}[tb]
 \centerline{\includegraphics[width=8.5cm]{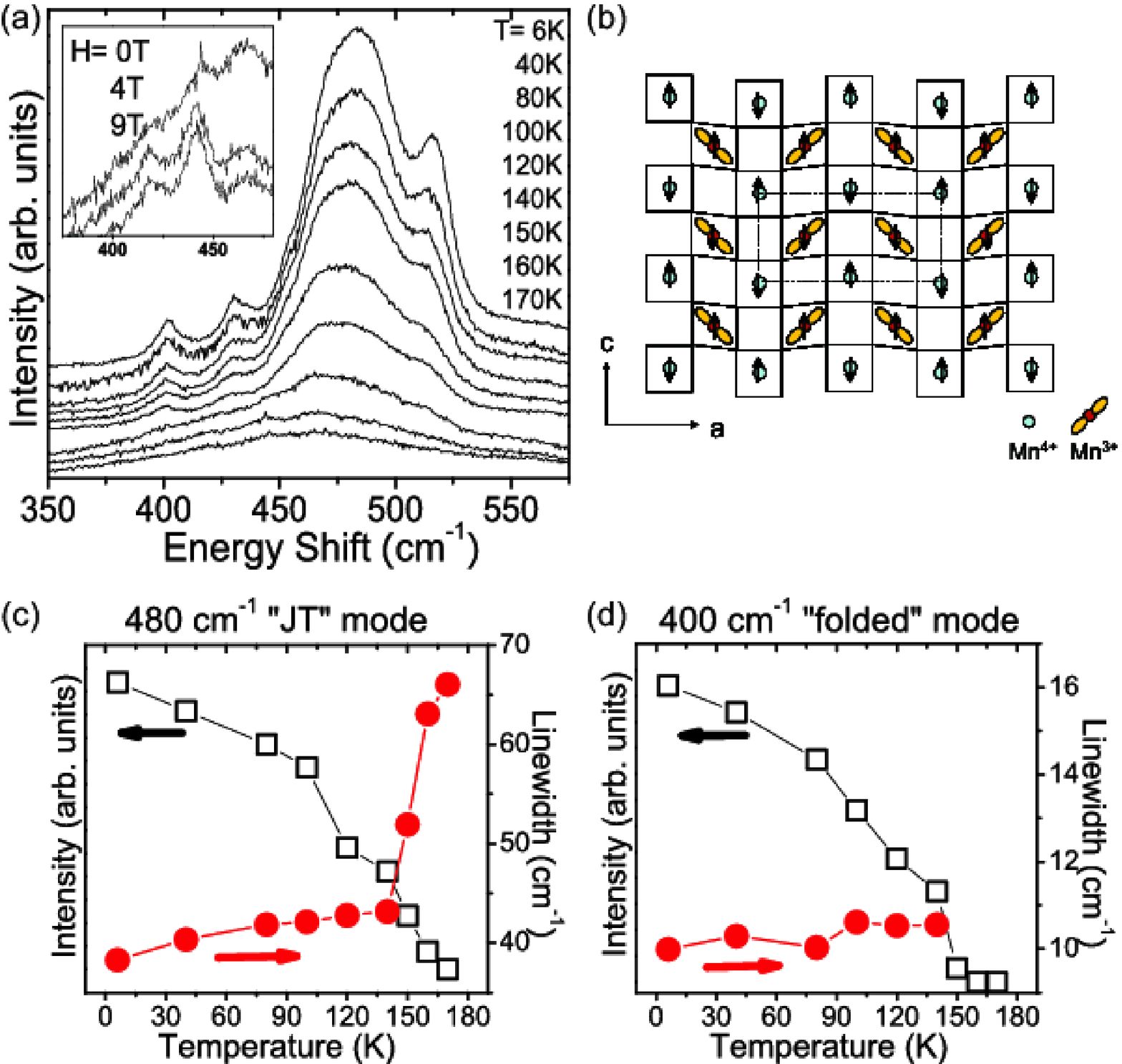}}
 \caption{\label{fig1}(a) Temperature dependence of the Raman spectrum
  of {\lcmo} in the temperature range 6--170 K.  (inset) Field-dependence
  of the 420 {\icm} and 440 {\icm}  ``field-induced modes'' for T=160 K,
  showing the enhancement of these modes with applied magnetic field.  (b) Schematic representation of the $CE$-type AFM/charge/orbital order
  in {\lcmo}, showing the doubling of the unit cell.  (c) Intensity
  (empty squares) and linewidth (filled circles) of the 480 {\icm}
  JT mode as a function of temperature.  The intensity of this mode is suppressed by $\sim$85\% from $T$=6 K to $T$=170 K.  (d) Summary of the intensity
  (empty squares) and linewidth (filled circles) of the 400 {\icm}
  ``folded phonon mode'' as a function of temperature.  The intensity of this mode is completely suppressed above $T$=150 K.  The symbol sizes in parts (c) and (d) reflect an estimated 5\% error associated with fits to the observed spectra.}
\end{figure}

\section{Results and Discussion}
\subsection{\label{data:mode}Mode assignments}
The temperature dependent Raman spectra of
La$_{0.5}$Ca$_{0.5}$MnO$_3$ are shown in Figure 1(a).  As the
temperature is decreased below $T_N$=180 K, a broad peak near 465
cm$^{-1}$ narrows, increases in intensity, and shifts to higher
energies, attaining a value of roughly 480 cm$^{-1}$ below
T$\sim$150 K.  The evolution of this peak is concomitant with the
growth of new peaks at 400, 430, and 515 cm$^{-1}$ that appear in
the COO phase.  Notably, the temperature dependent Raman spectra we
observe in Fig. 1(a) are consistent with those reported previously
by Granado et al.,~\cite{granado98} Liarokapis et
al.,~\cite{liarokapis99} and Abrashev et al.~\cite{abrashev01}  
Also, a full symmetry analysis of the Raman spectra in both LL and
 LR geometries was previously reported in Naler et al.[\onlinecite{naler01}] 
In the following, we focus on the assignments of several phonon
modes---including the 480 cm$^{-1}$ phonon and newly observed modes
at 400, 430, and 515 cm$^{-1}$ in the COO phase---which are used in
this study to monitor the field- and temperature-dependent evolution
of various phases in La$_{0.5}$Ca$_{0.5}$MnO$_3$ and
La$_{0.25}$Pr$_{0.375}$Ca$_{0.375}$MnO$_{3}$:

\begin{figure}[tb]
 \centerline{\includegraphics[width=8.5cm]{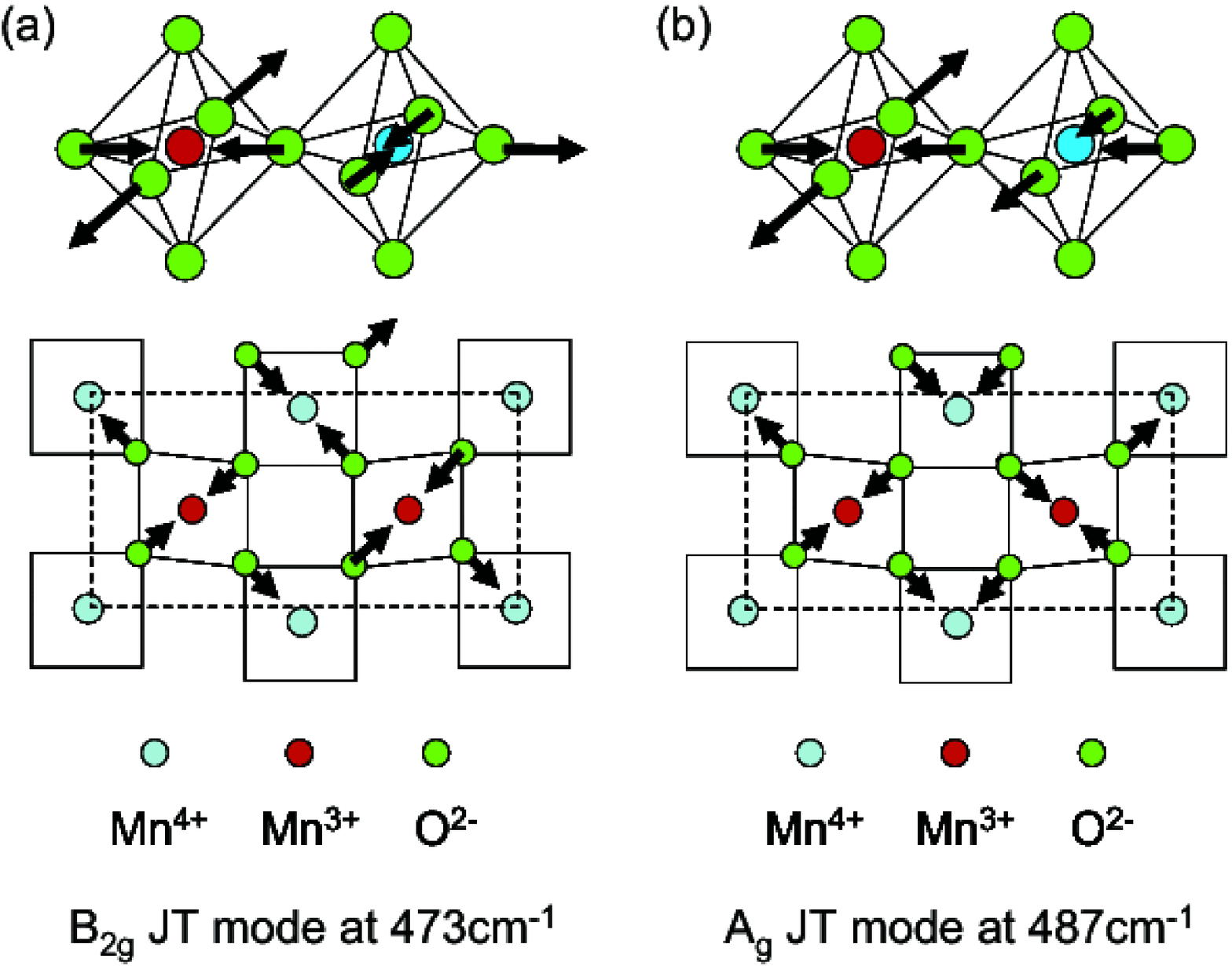}}
 \caption{\label{fig2}Normal modes associated with the 480 {\icm} JT peak:
   the 487 {\icm} $B_{2g}$ and 473 {\icm} $A_g$ asymmetric stretch modes
 of the oxygen atoms in the MnO$_6$ octahedra. (adapted from
 Ref.~\onlinecite{abrashev01}
 and Ref.~\onlinecite{yamamoto00})}
\end{figure}

\vspace{\baselineskip} (i) \textit{480 cm$^{-1}$ ``Jahn-Teller''
mode} - The 480 cm$^{-1}$ Jahn-Teller mode appears in both the LL and LR geometries.[\onlinecite{naler01}]  
Previous reports have attributed the 480 cm$^{-1}$ peak to
either Mn-O bending~\cite{abrashev01,iliev98} or
stretching~\cite{yamamoto00,liarokapis99,martin02} modes associated
with the MnO$_6$ octahedra.  We associate the 480 cm$^{-1}$ peak
with two nearly degenerate asymmetric Mn-O stretching modes of the
MnO$_6$ octahedra (illustrated in Fig. 2), a $B_{2g}$ mode at 473
cm$^{-1}$ and an $A_g$ mode at 487 cm$^{-1}$, based upon the
following evidence:  First, Raman studies show that the energy of
the 480 cm$^{-1}$ phonon is not sensitive to substitution at the
rare-earth site, $R$.~\cite{martin02}  This suggests that the 480
cm$^{-1}$ phonon in La$_{0.5}$Ca$_{0.5}$MnO$_3$ is a Mn-O stretch
mode, since substitution on the rare earth site R strongly affects
the frequencies of the Mn-O bending modes.~\footnote{The frequencies
of stretching modes depend on the Mn-O bond length while those of
bending modes depend on the $R$-O bond length in $R$MnO$_3$, and the
Mn-O bond length doesn't significantly change depending the $R$
element type but the $R$-O bond length does.  See
Ref.~\onlinecite{martin02}.}  Second, the broad band near 465
cm$^{-1}$---from which the 480 cm$^{-1}$ phonon evolves---is
activated in the Raman spectrum by JT distortions of the MnO$_6$
octahedra,~\cite{abrashev01,martin02,iliev01,abrashev99} which lower
the site symmetry and activate the asymmetric stretching modes.
Consequently, the 480 cm$^{-1}$ ``Jahn-Teller'' mode serves as an
ideal probe of the degree to which JT distortions evolve through
various temperature- and field-dependent transitions in
La$_{0.5}$Ca$_{0.5}$MnO$_3$ and
La$_{0.25}$Pr$_{0.375}$Ca$_{0.375}$MnO$_{3}$.  The JT distortions of
the MnO$_6$ octahedra are induced by the localization of $e_g$
electrons on the Mn$^{3+}$ sites, and consequently the presence of
the 480 cm$^{-1}$ mode can indicate even short-range charge/orbital
order.

\begin{figure}[tb]
 \centerline{\includegraphics[width=8.5cm]{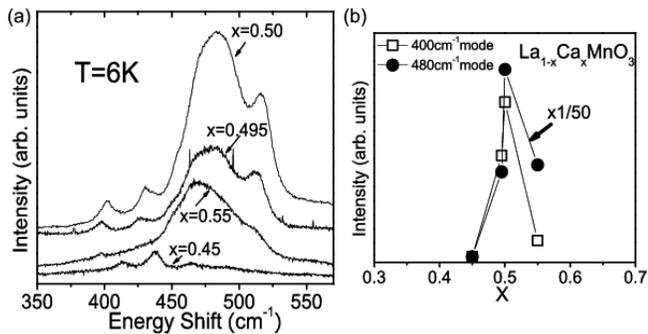}}
 \caption{\label{fig3}(a) Doping dependence of the Raman spectrum of La$_{1-x}$Ca$_x$MnO$_3$
 at $x$=0.5, 0.495, 0.55, and 0.45.  (b) Summary of the intensity of the 400 {\icm} ``folded phonon'' mode
 (empty squares) and the 480 {\icm} JT mode (filled circles) as a function of doping.}
\end{figure}

\vspace{\baselineskip} (ii) \textit{``Folded phonon modes''} - The
appearance of new peaks at 400, 430, and 515 cm$^{-1}$ in the COO
phase of La$_{0.5}$Ca$_{0.5}$MnO$_3$ below $T$$\sim$150 K (see Fig.
1(a)) reflects the doubling of the unit cell in the CO phase, which
``folds'' zone-boundary phonon modes to the zone center.  Similar
zone-folding behavior has been observed in the Raman spectra of
other COO materials, including La$_{0.5}$Sr$_{1.5}$MnO$_{4}$ and
LaSr$_2$Mn$_2$O$_7$.~\cite{yamamoto00}$^{,}$~\footnote{This Raman
activation due to the symmetry lowering is well described with the
$A_g$ JT mode in Fig 2(b).  Therefore, the $A_g$ JT mode, in fact,
gives information about the symmetry lowering to $P2_1/m$ as well as
the degree of JT distortions.}  The modes at 400 cm$^{-1}$ and 430 cm$^{-1}$
appear in both the LL and LR geometries, and the mode at 515 cm$^{-1}$ 
appears in the LR geometry.[\onlinecite{naler01}]  We
attribute the new $B_{2g}$ modes at 400 cm$^{-1}$ and 430 cm$^{-1}$,
and the $A_g$ mode at 515 cm$^{-1}$, to Mn-O bending modes of the
MnO$_6$ octahedra,~\footnote{We've already assigned the 480 {\icm}
peak to the JT modes, and the observed frequencies, 400 {\icm} and
430 {\icm}, correspond to the frequency dependence of bending modes
reported by Mart{\'i}n-Carr{\'o}n et al.~\cite{martin02}  The 515
{\icm} mode is already assigned to one of the bending modes by
Abrashev et al.~\cite{abrashev02}} which become Raman-active in the
monoclinic $P2_1/m$ structure that results from the doubling of the
\textit{Pnma} phase by charge/orbital
ordering.~\cite{abrashev01,iliev01}  Evidence for our interpretation
is seen in the rapidity with which these modes disappear as
La$_{1-x}$Ca$_x$MnO$_3$ is doped away from commensurate $x$=0.5
filling, as shown in Figure 3.  The strong doping dependence of
these `activated' modes is indicative of the sensitivity of these
folded modes to the loss of long-range charge and orbital order;
consequently, we can use the intensities of these folded phonons to
characterize the degree of long-range charge/orbital order in
La$_{0.5}$Ca$_{0.5}$MnO$_3$ as functions of magnetic field and
temperature.

\vspace{\baselineskip} (iii) \textit{``Field-induced modes''} - Two
modes at 420 cm$^{-1}$ and 440 cm$^{-1}$, which are very weak but
apparent in the zero-field spectrum at $T$=160 K, are dramatically
enhanced with increasing magnetic field, as shown in the inset of
Figure 1(a).  These magnetic-field-induced modes are characteristic
of spectra that have been observed for manganites having structural
phases with weak or no JT distortions, i.e., with symmetric or
nearly symmetric MnO$_6$ octahedra.  These phases include
rhombohedral La$_{0.67}$Sr$_{0.33}$MnO$_3$,~\cite{martin02}
orthorhombic (FMM)
La$_{0.67}$Ca$_{0.33}$MnO$_3$,~\cite{martin02,iliev01,abrashev99}
rhombohedral La$_{0.98}$Mn$_{0.96}$O$_3$,~\cite{iliev01} and
orthorhombic (PM) CaMnO$_3$.~\cite{abrashev02}  Additionally,
magnetic-field-dependent x-ray diffraction studies of
La$_{0.5}$Ca$_{0.5}$MnO$_3$ provide further evidence that the MnO$_6
$ octahedra have identical Mn-O bond lengths at high
fields.~\cite{tyson04}  Therefore, we assume that the appearance of
the new modes at 420 cm$^{-1}$ and 440 cm$^{-1}$ are indicative of a
new structural phase that has no, or only weak, JT distortions.

Regarding the specific mode assignment of the 420 cm$^{-1}$ and 440
cm$^{-1}$ phonons, although it is unclear whether these modes are
best associated with an orthorhombic $Pnma$ structure (of LaMnO$_3$)
or with the rhombohedral structure,~\cite{iliev01,abrashev99} there
is a clear correspondence between the 420 cm$^{-1}$ and 440
cm$^{-1}$ field-induced modes of La$_{0.5}$Ca$_{0.5}$MnO$_3$ and the
465cm$^{-1}$ and 487cm$^{-1}$ modes in CaMnO$_3$;~\cite{abrashev02}
the $\sim$46 cm$^{-1}$ frequency difference between these pairs of
modes can be attributed to the strong sensitivity of the frequency
of these phonons to the R-O bond length reported by
Mart{\'i}n-Carr{\'o}n et al.~\cite{martin02}  In CaMnO$_3$, the
modes at 465 cm$^{-1}$ and 487 cm$^{-1}$ are activated by a
$D_{[101]}$ type distortion, which involves a rotation (tilt)
of the MnO$_6$ octahedra about the [101] cubic axis.  This distortion is
imposed on the cubic perovskite structure by the $Pm\overline{3}m$
symmetry, and results in a lowering of symmetry to the
orthorhombic \textit{Imma} structure.~\cite{abrashev02}  However,
because CaMnO$_3$ also has both weak JT distortions and
$D_{A-shift}$ type distortions of the MnO$_6$ octahedra, CaMnO$_3$ is most
appropriately associated with the orthorhombic $Pnma$
structure.~\cite{abrashev02}  The latter distortion involves a shift of the
La/Ca atoms from their sites in the ideal perovskite in [100] direction
(in the Pnma basis).  Therefore, more study is needed to
identify whether the specific structural phase we observe in
La$_{0.5}$Ca$_{0.5}$MnO$_3$ at high magnetic fields is a $Pnma$
phase with weak JT distortions, or simply a $Imma$ phase with a
$D_{[101]}$ basic distortion. In either case, we can use the
field-induced modes at 420 cm$^{-1}$ and 440 cm$^{-1}$ to monitor
the evolution of a ``weakly JT distorted'' ($Imma$ or $Pnma$) phase
in La$_{0.5}$Ca$_{0.5}$MnO$_3$ and
La$_{0.25}$Pr$_{0.375}$Ca$_{0.375}$MnO$_{3}$ as functions of
magnetic field and temperature.

\subsection{Raman studies of La$_{0.5}$Ca$_{0.5}$MnO$_3$}
In the following, we describe the results of our temperature- and
field-dependent Raman scattering measurements of
La$_{0.5}$Ca$_{0.5}$MnO$_3$, in which material the COO state is
fully commensurate and ordered (see Fig. 1(b)) at temperatures well
below $T_{CO}$=240 K.

\subsubsection{Temperature dependence}
Fig. 1(a) shows that the four low temperature peaks at 480
cm$^{-1}$, 400 cm$^{-1}$, 430 cm$^{-1}$, and 515 cm$^{-1}$
systematically lose intensity as a function of increasing
temperature:  the ``folded phonon'' peaks at 400 cm$^{-1}$, 430
cm$^{-1}$, and 515 cm$^{-1}$ are completely suppressed above
$T$$\sim$150 K, but exhibit very little change in linewidth
throughout the temperature range 0--150 K.  On the other hand, the
480 cm$^{-1}$ ``JT mode'' persists above 150 K, but exhibits both a
significant broadening in linewidth and a shift in energy above this
temperature.  The temperature dependent intensities and linewidths
of the 480 cm$^{-1}$ and 400 cm$^{-1}$ modes are summarized in
Figures 1(c) and 1(d).  Fig. 1(a) also reveals that new weak peaks
at 420 cm$^{-1}$ and 440 cm$^{-1}$ start to appear at $T$$\sim$150
K, above which temperature the folded phonon peaks are suppressed
and the 480 cm$^{-1}$ JT mode is weak but present.  As shown in the
inset of Fig. 1(a), the new peaks at 420 cm$^{-1}$ and 440 cm$^{-1}$
are very weak in the $H$=0 T spectra, but become significantly
stronger with increasing magnetic field.

\begin{figure}[tb]
 \centerline{\includegraphics[width=7.5cm]{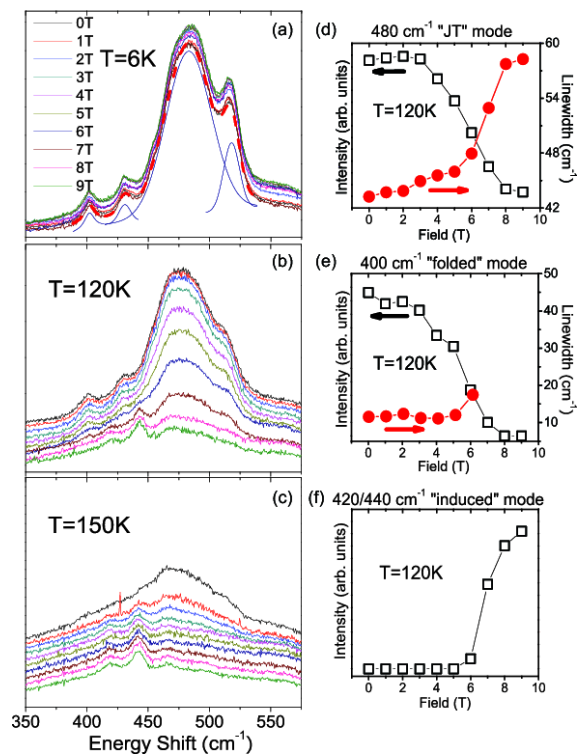}}
 \caption{\label{fig4}(a) Magnetic field dependence of the Raman spectrum of {\lcmo} at 6 K,
 illustrating the absence of a significant change in the spectrum with increasing magnetic field.
 The solid lines illustrate the examples of Gaussian and Lorentzian fits to the actual spectra, and the thick dashed line illustrates the sum of all individual fits.
 (b) Magnetic field dependence of the Raman spectrum of {\lcmo} at 120 K, showing a significant
 suppression of the JT mode and the folded phonon modes, and indicating the appearance
 of field-induced modes near 420 {\icm} and 440 {\icm} with increasing magnetic field.
 (c) Magnetic field dependence of the Raman spectrum of {\lcmo} at 150 K, showing the evolution
 of the 420 {\icm} and 440 {\icm} field-induced modes with increasing field.
 (d) Summary of the intensity (empty squares) and linewidth (filled circles)
 of the 480 {\icm} ``JT mode'' as a function of magnetic field.  The intensity of this mode is suppressed by $\sim$80\% from $H$=0 T to $H$=9 T.
 (e) Summary of the intensity (empty squares) and linewidth (filled circles)
 of the 400 {\icm} ``folded phonon'' mode as a function of magnetic
 field.  The intensity of this mode is completely suppressed above $H$=8 T.
 (f) Summary of the combined intensities (empty squares)
 of the 420/440 {\icm} ``field-induced phonon'' modes as a function of magnetic field.  The intensity of these modes is essentially fully developed above $H$=8 T.  The symbol sizes in parts (d), (e), and (f) reflect an estimated 5\% error associated with fits to the observed spectra.}
\end{figure}

\subsubsection{\label{data:lcmo:field}Field dependence}
Figs. 4(a), 4(b), and 4(c) show the field-dependent Raman spectra
obtained at $T$=6 K, 120 K, and 150 K, respectively.  Below $T$=80
K, Raman spectra taken as a function of magnetic field show little
significant change in the linewidths or intensities of the modes
between 0 T to 9 T, as shown in the spectra obtained at $T$=6 K in
Fig. 4(a).  However, for $T$$\geq$80 K, the intensities of the
spectra become increasingly sensitive to the application of a
magnetic field, as shown in the spectra obtained at $T$=120 K in
Fig. 4(b).
 Figs. 4(d) and 4(e) summarize the field-dependent intensities and
linewidths of both the 480 {\icm} ``JT mode'' and the 400 {\icm}
``folded phonon mode'' at $T$=120 K.
The parameters of the 480 {\icm} ``JT mode'' were obtained from Gaussian fits and those of the 400 {\icm} ``folded phonon mode'' were obtained from Lorentzian fits to the spectra in Fig. 4, examples of which are illustrated by the solid curves in Fig. 4 (a).~\footnote{The 480 {\icm} ``JT mode'' was better fit to the Gaussian curve than the Lorentzian curve since it's composed of two closely adjacent modes.}  The symbol sizes in Figs. 4(d) and (e) reflect the estimated 5\% errors associated with these fits.
The field-dependent development of both the intensities and linewidths of these peaks is
very similar to their temperature-dependent evolution described
earlier.
 In particular, while there are significant and systematic changes in
the intensities of the ``JT'' and ``folded phonon'' modes with
increasing field for $H$$<$6 T, there are much less dramatic changes in the
linewidths of these modes with applied fields.  However, the linewidths also show significant changes for 6 T$\leq$$H$$<$8 T. For $H$$\geq$8 T,
the folded phonon mode disappears, but the 480 {\icm} JT mode persists with a residual intensity.  The spectrum
at $T$=120 K also illustrates that the new modes at 420 {\icm} and
440 {\icm} appear at $H$$\sim$6 T, indicating the appearance of a ``weakly
JT distorted'' phase.  These modes become enhanced with increasing
field but show no significant enhancement For $H$$\geq$8 T, indicating a
full development of the modes.  The spectrum at $T$=150 K in Fig.
4(c) shows that the evolution of the ``field-induced'' modes occurs
at a significantly lower magnetic field value $H^*(T)$ compared to
that at $T$=120 K, and the modes are completely developed even at
$H$$\sim$3 T.
 One also observes that, at $T$=150 K, the folded phonon modes are very
weak at $H$=0 T, and disappear with even the smallest applied
magnetic field, suggesting that long range COO is disrupted at this
temperature even for small applied magnetic fields.  We show the
evolution of the folded phonon modes and the field-induced modes by
plotting their intensities as functions of both magnetic field and
temperature in Figs. 5(a) and 5(b).

\subsubsection{Discussion}
The temperature-dependent Raman spectra for {\lcmo} allow us to draw
several conclusions regarding the manner in which COO in {\lcmo}
melts with increasing temperature.  First, the persistence with
increasing temperature of the folded-phonon modes at 400, 430, and
515 {\icm}---which are again highly sensitive to the degree of
long-range COO---suggest that long-range coherence of COO is
preserved throughout much of the thermal melting process.  This
conclusion is also supported by the temperature dependence of the
480 {\icm} and 400 {\icm} mode linewidths:  in the temperature
regime well below the transition temperature ($T$$<$150 K in Fig.
1(c) and (d))---in which the size of the COO domain regions is much
larger than the mean free path of the phonon---the 480 {\icm} and
400 {\icm} mode linewidths show no dependence on changing
temperature.  However, in the coexistence regime around the
transition temperature ($T$$>$150 K in Fig. 1(c) and (d)), the
480{\icm} mode linewidth exhibits a rapid increase with temperature,
indicating that the sizes of the COO domain regions have become
smaller than the phonon mean free path.  This temperature-dependent
evolution is suggestive of a first-order ``electronic
crystallization'' transition in {\lcmo}, in which the kinetics in
the coexistence region corresponds to the propagation of a phase
front.  Indeed, our evidence that thermal melting in {\lcmo}
proceeds heterogeneously, by disrupting order at the surfaces of COO
domain regions rather than by homogeneously disrupting COO domains,
is consistent with TEM studies showing heterogeneous melting of COO
in La$_{0.33}$Ca$_{0.67}$MnO$_3$.~\cite{tao04}  Furthermore, the
first-order nature of the COO melting transition in
three-dimensional (3D) {\lcmo} is also supported by the fact that
this transition differs substantially from the continuous
(second-order) CDW transition in layered (quasi-2D) dichalcogenides
such as $1T$-TiSe$_2$:  in the latter system, there is roughly a
five-fold increase in the linewidth of the CDW amplitude modes with
increasing temperature or pressure, suggesting a very rapid---and
roughly uniform---dissolution of long-range CDW order throughout the
melting process.~\cite{snow03}  It is interesting to note that the
observed differences in the ordering transitions of 3D {\lcmo} and
quasi-2D $1T$-TiSe$_2$ are also consistent with theoretical
calculations of ordering behavior in 3D and 2D systems:  variational
renormalization group calculations of the three-dimensional
three-state Potts model---which is an appropriate description of the
three-fold degeneracy associated with ordering in cubic
{\lcmo}---predict that the ordering transition should be weakly
first order.~\cite{Nienhuis81,blote79}  The addition of lattice
degrees of freedom (i.e., a compressible system) has been shown to
further ``harden'' the first-order nature of the transition in 3D
systems.~\cite{baker71,bergman76}  On the other hand, calculations
of the two-dimensional three-state Potts model---which is
appropriate for layered dichalcogenides such as $1T$-TiSe$_2$
because of the three commensurate CDWs in these quasi-2D
materials---predict that the ordering transition should be
continuous.~\cite{Nienhuis81}  We note, however, that the presence
of a critical amount of disorder in 3D systems is expected to round
the first order transition and make the transition difficult to
distinguish from a continuous phase
transition.~\cite{imry79,aizenman89}

Our results allow us to identify several distinct electronic and
structural phases in cubic {\lcmo} as a function of temperature: (i)
\textit{Long-range COO regime} - Below 150 K, long-range COO
persists in {\lcmo}, as suggested by the presence of folded phonons
and by the temperature-insensitive linewidths of the 480 {\icm} JT
mode.  (ii) \textit{Short-range COO regime} - In the temperature
range 150--170 K, long-range coherence of COO in {\lcmo} becomes
suppressed---as evidenced in this temperature regime by the
persistence of a 480 {\icm} JT mode with a significantly broadened
linewidth, and by the disappearance of the folded phonon
modes---indicating a phase regime in which short-range COO prevails.
The appearance of new modes at 420 {\icm} and 440 {\icm} in this
temperature range, 150$\leq$$T$$\leq$170 K, also indicates that
``weakly JT distorted'' phase regions coexist with the strongly JT
distorted COO regions in this regime.  (iii)\textit{ Weakly JT
distorted regime} - At very high temperatures ($T$$>$200 K), the 480
{\icm} JT mode is almost completely suppressed, and the 420 {\icm}
and 440 {\icm} modes are clearly evident, indicating the presence of
a weakly JT-distorted FMM regime associated with either a $Imma$ or
$Pnma$ phase in this temperature range.

These results suggest the following description of the thermal
melting process in {\lcmo}:  With increasing temperature below 150
K, first-order melting primarily proceeds at the interface between
large coherent COO regions, thereby shrinking the COO volume, but
maintaining long-range coherence of COO domains; at still higher
temperatures (150$\leq$$T$$\leq$170 K), additional melting leads to
the eventual collapse of long-range COO, and the development of a
regime in which short-range COO coexists with a ``weakly JT
distorted'' phase.  This melting process is consistent with the
thermal melting of COO observed in earlier studies of {\lcmo}, in
which a commensurate (CM) charge/orbital ordered phase was observed
to melt with increasing temperature into an incommensurate (IC)
charge ordering regime that coexists with the ferromagnetic metal
phase above $T_N$=180K.~\cite{radaelli95,radaelli9755,chen96}  In
fact, the high temperature modes we observe at 420 {\icm} and 440
{\icm}, which are indicative of the weakly JT distorted phase, are
known to be associated with the Raman spectrum of ferromagnetic
metallic La$_{0.55}$Ca$_{0.45}$MnO$_3$ shown in Figure 3(a).  Our
description of the thermal melting process in {\lcmo} is also
consistent with the three-step ``heterogeneous'' formation process
of charge ordering (CO) observed by Tao and Zuo in
La$_{0.33}$Ca$_{0.67}$MnO$_3$ using transmission electron microscopy
(TEM):~\cite{tao04}  short-range CO in La$_{0.33}$Ca$_{0.67}$MnO$_3$
initially forms at T$\sim$280K; CO regions then grow with decreasing
temperature via cluster formation; and finally, long-range CM CO
develops via the percolation of these clusters below
$T$$\sim$235K.~\cite{tao04}  Because Raman scattering is a q=0 probe, we are not able to estimate the size
of the CO domains. However, previous electron microscopy studies
have estimated their size to be approximately 10-20 nm at 200 K and 50-60 nm at 125 K.~\cite{mori98}

Interestingly, our field-dependent Raman results in {\lcmo} (see
Sec.~\ref{data:lcmo:field}) indicate a strong similarity between the
thermal and field-induced disruption of COO in this material. For
example, the persistence of the folded phonon modes---as well as the
insensitivity of the 480 {\icm} and 400 {\icm} mode
linewidths---with increasing magnetic field below 6 T at $T$=120 K
suggests that domains characterized by long-range COO persist
throughout much of the field-induced melting process.  Second, the
appearance of new field-induced modes, and the persistence of the
480 {\icm} JT mode above 6 T at $T$=120 K, are indicative of the
coexistence in this high field range of a weakly JT distorted/FMM
phase (associated with either a $Imma$ or $Pnma$ structure; see
Sec.~\ref{data:mode}iii) and a short-range COO phase.

\begin{figure}[tb]
 \centerline{\includegraphics[width=5cm]{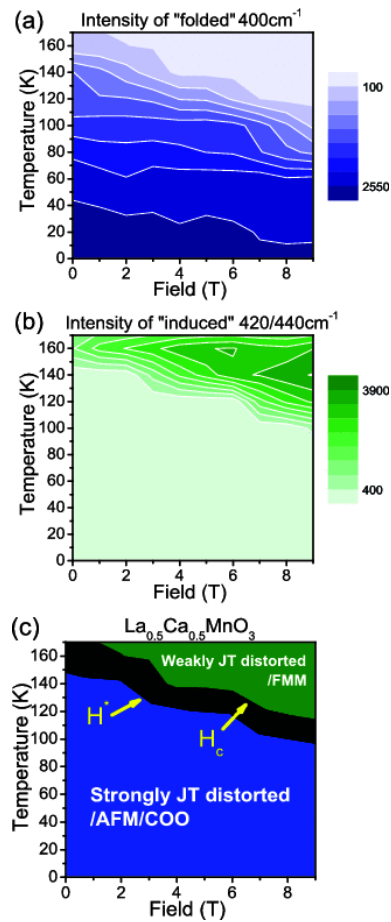}}
 \caption{\label{fig5}(a) Contour plot of the intensity of the 400 {\icm} folded phonon mode
 as functions of temperature and magnetic field.  The dark (purple) shades
 indicate the prevalence of the strongly JT distorted ($P2_1/m$ space group) structural phase,
 which is associated with the AFM/COO phase.
 (b) Contour plot of the combined intensities of the 420 {\icm} and 440 {\icm} field-induced modes
 as functions of temperature and field.  The dark (green) shades
 indicate the prevalence of the weakly JT distorted ($Imma$ or $Pnma$ space group) structural phase,
 which is associated with the ferromagnetic metal (FMM) phase.
 (c) Temperature/magnetic field phase diagram of {\lcmo} inferred from the intensity plots in (a) and (b),
 showing the phase boundary, $H^*(T)$, between the strongly JT distorted ($P2_1/m$ space group) AFM/COO phase (purple)
 and intermediate ``coexistence'' phase (black) regimes, and the phase boundary,
 $H_c(T)$, between the intermediate ``coexistence'' phase (black)
 and weakly JT distorted ($Imma$ or $Pnma$ space group) FMM phase (green) regimes.}
\end{figure}

The similarities between thermal and field-induced melting of COO in
{\lcmo} suggest that the application of a magnetic field also
disrupts antiferromagnetic order preferentially at the surface of
COO regions, giving rise to heterogeneous melting of the coherent
long-range COO regions, and the evolution of a field regime in which
short-range COO regions coexist with weakly JT distorted/FMM
regions.  Notably, our field-dependent Raman results suggest a
three-stage field-dependent melting process in {\lcmo}, consistent
with the three distinct structural phases identified by Tyson et
al.~\cite{tyson04} in field-dependent x-ray absorption measurements
at $T$=115 K.  Indeed, although the structural details inferred by
the x-ray measurements were obtained only at $T$=115
K,~\cite{tyson04} the comparisons of these x-ray results to our more
extensive temperature- and field-dependent Raman measurements
suggest that one can generalize the following structural details
throughout $H$-$T$ phase diagram:  (i) \textit{Low field-regime} -
At low magnetic fields (e.g., 0$\leq$$H$$\leq$6 T at $T$=120 K),
only the 480 {\icm} ``JT mode'' and the 400 {\icm} ``folded phonon''
modes are present, indicative of a long-range ``strongly JT
distorted''/COO phase.
 X-ray absorption measurements (at $T$=115 K) confirm that the
distribution of Mn-O and Mn-Mn separations exhibit only a weak field
dependence in this regime.~\cite{tyson04}  (ii) \textit{High field
regime} - Above a critical field value $H_c(T)$ (e.g., $H_c$$\sim$7
T at $T$=120 K), the 400 {\icm} folded phonon mode is completely
suppressed, and the field-induced 420 and 440 {\icm} modes are fully
developed, indicative of a long-range ``weakly JT distorted''/FMM
phase associated with either a $Imma$ or a $Pnma$ structure.
 Field-dependent x-ray data at $T$=115 K indicate that the
transformation to this high-field FM regime is accompanied, first,
by a reduction of the Debye-Waller (DW) factor for the Mn-O
distribution to ${\sigma}^2{\sim}$0.0032, which is comparable to
that of CaMnO$_3$, and second, by the development of a Gaussian
radial Mn-O distribution, indicating the disappearance of JT
distortions. The average Mn-O bond distance also approaches a value
of 1.96 {\AA} in the high-field FM region, which corresponds to the
average Mn-O bond distance found in FM La$_{1-x}$Ca$_x$MnO$_3$ near
$x$=0.33.~\cite{tyson04}  (iii) \textit{Intermediate field regime} -
In the intermediate field range (e.g., 6$\leq$$H$$\leq$7 T at
$T$=120 K), the folded phonons are very weak, and the 420 and 440
{\icm} modes start to appear, indicative of a coexistent phase
regime with a combination of strongly JT distorted/COO and weakly JT
distorted/FMM phase regions.  A narrow mixed-phase region was also
observed at intermediate field-values by x-ray absorption
measurements.~\cite{tyson04}  The similarity between thermal and
field-induced melting behavior of the 480 {\icm} ``JT mode'' in
{\lcmo} suggests that---like the $T$-dependent phase
transition---the field-induced COO transition is a first-order
quantum phase transition associated with melting of an electronic
``crystal,'' and that the ``intermediate field regime'' can be
identified with the coexistence phase regime in which the phonon
mean free path is on the order of, or longer than, the size of the
COO domain regions.

The three phase regimes implied by our Raman results are summarized
in the magnetic-field- and temperature-dependent phase diagram of
{\lcmo}, shown in Figure 5(c), obtained by combining the intensity
plots of the folded phonon (Fig. 5(a)) and field-induced (Fig. 5(b))
modes.  The resulting structural $H$-$T$ phase diagram is indicative
of a very robust COO state in {\lcmo}, with critical field values,
$H^*(T)$---between the long-range COO and intermediate
``coexistence'' regimes---and $H_c(T)$---between the intermediate
``coexistence'' and weakly JT distorted FMM regimes---that decrease
roughly linearly with decreasing temperature at a rate of
$\sim$$-$0.2 T/K.
 Extrapolating this first-order critical field line to higher field
values suggests an approximate $T$=0 critical point of
$H^*(T$=$0)$$\sim$30 T for {\lcmo}.

\subsection{Raman studies of {\lpcmo}}
It is of interest to explore the effects of disorder on thermal and
field-induced melting of the robust COO state observed {\lcmo}, in
order to explore the manner in which disorder destabilizes COO in
the maganites.  The effects of disorder on commensurate COO in
{\lcmo} can be investigated by substituting Pr in the
La$_{1-x}$Ca$_x$MnO$_3$ lattice.  The random substitution by Pr ions
having a 3+ valence state results in the deviation from a
commensurate composition $x$=0.5, but preserves the COO state even
at the DE (double exchange)-optimized composition $x$=0.375, due to
the relatively small ionic size of Pr.~\cite{uehara99} Accordingly,
Pr-substitution into commensurate {\lcmo} has the effect of
disordering the commensurate charge-ordered lattice, as
schematically illustrated in Fig. 6(b).  In the following, we
describe the results of temperature- and field-dependent Raman
scattering measurements of {\lpcmo}, in which we investigate the
effects of disorder on temperature- and field-induced melting of
COO.

\begin{figure}[tb]
 \centerline{\includegraphics[width=8.5cm]{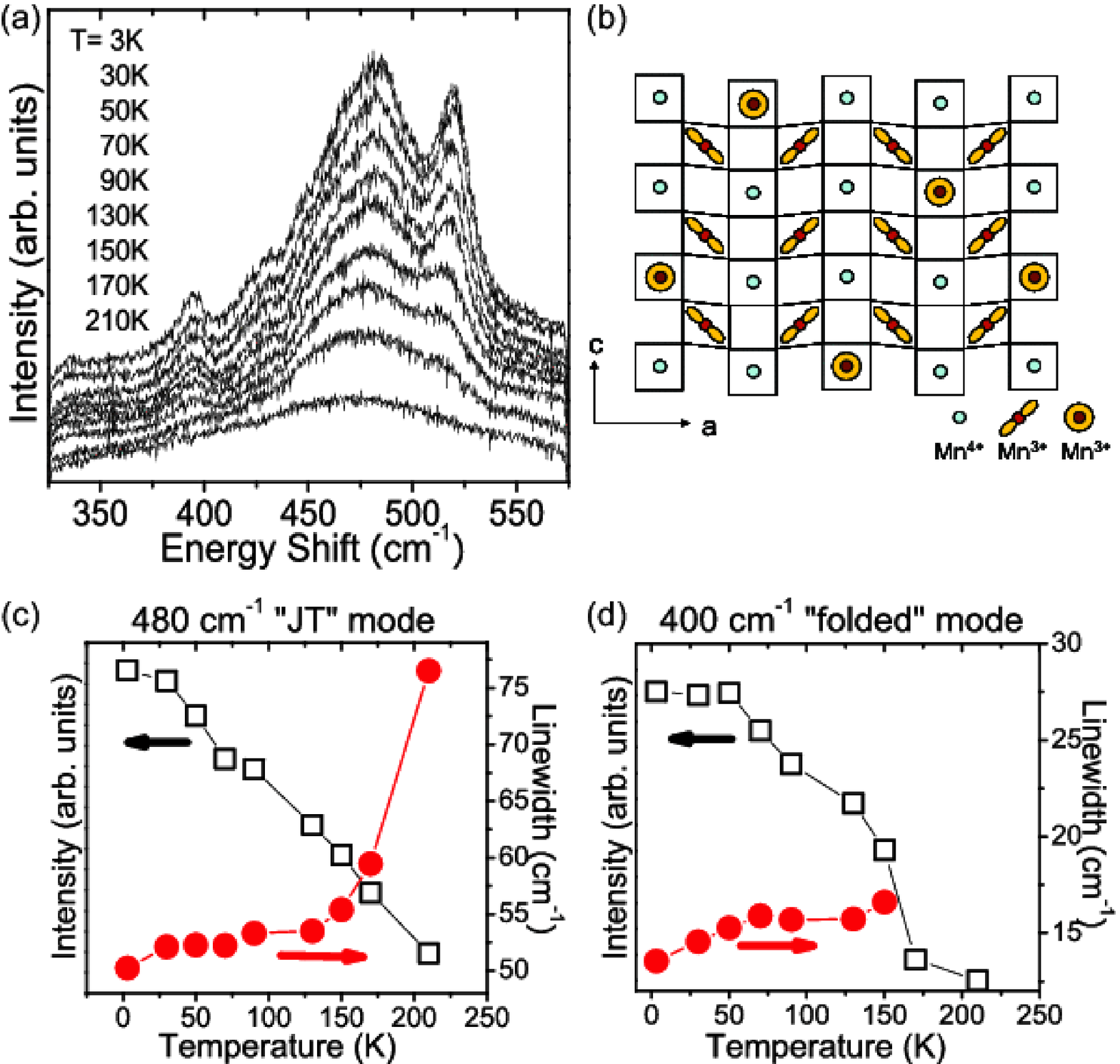}}
 \caption{\label{fig6}(a) Temperature dependence of the Raman spectrum of {\lpcmo} in the temperature range 3.5--210 K.
 (b) Schematic representation of the effect of Pr-substitution in {\lpcmo} on COO, where double circles represent charge (hole) defects in COO.
 (c) Summary of the intensity (empty squares) and linewidth (filled circles)
 of the 480 {\icm} ``JT mode'' of {\lpcmo} as a function of temperature.  The intensity of this mode is suppress by $\sim$75\% from $T$=3.5 K to $T$=210 K.
 (d) Summary of the intensity (empty squares) and linewidth (filled
 circles) of the 400 {\icm} ``folded phonon" mode of {\lpcmo} as a
 function of temperature.  The intensity of this mode is completely suppressed above $T$=170 T.  The symbol sizes in parts (c) and (d) reflect an estimated 5\% error associated with fits to the observed spectra.}
\end{figure}

\subsubsection{Temperature dependence}
The temperature dependent Raman spectra of {\lpcmo} are shown in
Figure 6(a).  Note that the temperature dependence of the Raman
spectra in the COO phase of {\lpcmo} is not significantly different
from that of {\lcmo}:  The ``folded modes'' at 400, 430, and 520
{\icm} gradually lose intensity with relatively little change in
linewidth, eventually becoming suppressed above $T$$\sim$170 K;
however, the 480{\icm} mode persists above $T$$\sim$170K, similar to
the temperature-dependent behavior observed in {\lcmo}.  The
remarkable similarity between the temperature dependences of the
{\lpcmo} and {\lcmo} Raman spectra suggest that disordering COO with
Pr-substitution does not appreciably affect temperature-dependent
melting of COO in this material.

\begin{figure}[tb]
 \centerline{\includegraphics[width=7.5cm]{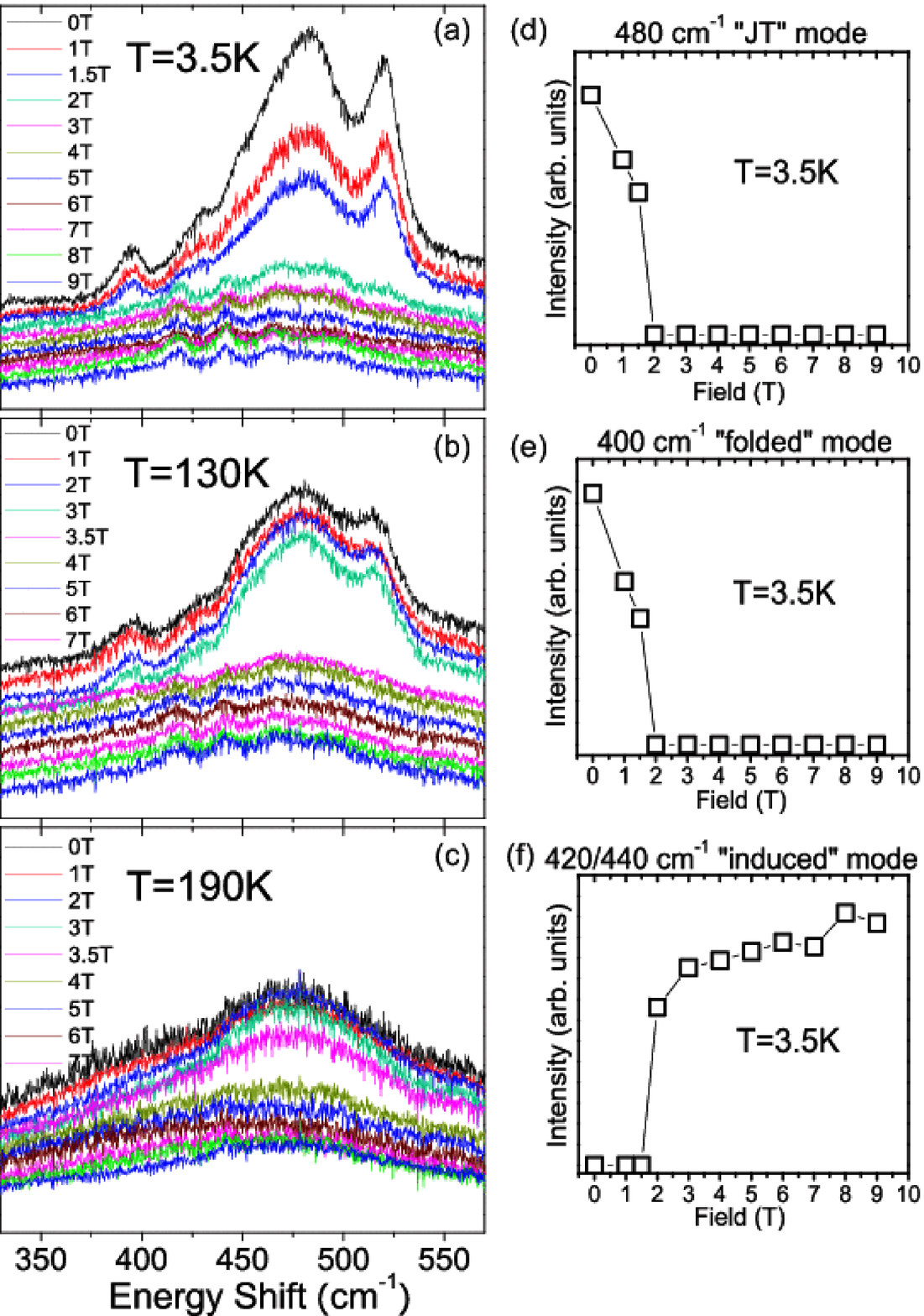}}
 \caption{\label{fig7}(a) Magnetic field dependence of the Raman spectrum of {\lpcmo} at 3.5 K,
 showing a dramatic suppression of the 480 {\icm} JT and 400 {\icm} folded phonon modes,
 and the appearance of the 420 {\icm} and 440 {\icm} field-induced modes at $H$$\sim$2 T.
 (b) Magnetic field dependence of the Raman spectrum of {\lpcmo} at 130 K,
 showing an abrupt field-induced transition between strongly JT distorted and weakly JT distorted structural phases.
 (c) Magnetic field dependence of the Raman spectrum of {\lpcmo} at 190 K.
 (d) Summary of the intensity (empty squares) and linewidth (filled circles)
 of the 480 {\icm} ``JT mode'' as a function of magnetic field.  The intensity of this mode is completely suppressed above $H$=2 T.
 (e) Summary of the intensity (empty squares) and linewidth (filled circles)
 of the 400 {\icm} ``folded phonon'' mode as a function of magnetic
 field.  The intensity of this mode is completely suppressed above $H$=2 T.
 (f) Summary of the combined intensities (empty squares)
 of the 420/440 {\icm} ``field-induced phonon'' modes as a function of magnetic field.  The intensity of this mode is essentially fully developed above $H$=2 T.  The symbol sizes in parts (d), (e), and (f) reflect an estimated 5\% error associated with fits to the observed spectra.}
\end{figure}

\subsubsection{Field dependence}
In contrast with the temperature-dependent evolution of the Raman
spectrum in {\lpcmo}, the field-dependent evolution of the {\lpcmo}
spectrum differs markedly from that observed in {\lcmo}, as
illustrated in Figs. 7.  Specifically, even in the lowest
temperature spectrum measured at $T$$\sim$3.5 K, shown in Fig 7(a),
the 480 {\icm} ``JT mode'' and the 400 {\icm} ``folded phonon mode''
of {\lpcmo} exhibit a dramatic and abrupt decrease in intensity with
increasing magnetic field, becoming suppressed above a critical
field of roughly $H_c$$\sim$2 T. Furthermore, above this critical
field, there is an abrupt appearance of the field-induced 420 and
440 {\icm} modes, signaling the appearance of a ``weakly JT''
distorted phase.  This strong field dependence is in strong contrast
to the field-independence observed in {\lcmo} for $T$$<$80 K.
Additionally, this strong field dependence is different than the
field dependence of {\lcmo} observed even in high temperatures
$T$$>$80 K. Specifically, {\lcmo} exhibits a gradual change in
intensities of all the modes with increasing field, as well as an
intermediate field regime in which both the folded phonon modes and
weakly JT distorted modes are observed, indicating the coexistence
of long-range COO and weakly JT distorted regions in this field
range. By contrast, the field dependence observed at $T$=3.5 K in
{\lpcmo} is much more dramatic, exhibiting an abrupt change in the
spectrum between 1.5 and 2 T without any evidence for an
intermediate field regime characterized by phase coexistence.
Interestingly, the abrupt change in the spectrum is even more
remarkable at higher temperatures, as illustrated by the field
dependent data at $T$=130 K shown in Fig. 7(b) and $T$=190 K shown in Fig. 7(c):  Specifically, at
$T$=130 K, the ``low-field regime'' spectrum shows no field
dependence up to $H$$\sim$3 T.  However,
there is a striking transition to the ``weakly JT distorted''
spectral response between 3 and 3.5 T.  At $T$=190 K, such a dramatic transition occurs between 3.5 and 4 T,
again with no evidence for an intermediate field regime characterized by phase coexistence.  Above
the critical field, the ``weakly JT distorted'' spectrum shows no
additional field dependence; that is, this field-induced transition
in {\lpcmo} appears to be an abrupt and direct change from the
COO/AFM phase to the weakly JT distorted/FM phase.

Figure 8(a) and 8(b) illustrate the field dependence of the
intensities of the 400 {\icm} ``folded phonon'' and the 420 and 440
{\icm} field-induced modes in {\lpcmo} in the temperature range
3.5--210 K.  These intensity plots enable us to map out a structural
and COO phase diagram for {\lpcmo} in the temperature range 3.5--210
K and in the field range 0--9 T.  Interestingly, unlike {\lcmo}, the
critical field $H_c$ between strongly JT distorted/COO and weakly JT
distorted/FMM phases in {\lpcmo} doesn't show a monotonic decrease
as a function of increasing temperature, but rather increases
systematically from $\sim$2 T to 4 T with increasing temperature.

\begin{figure}[tb]
 \centerline{\includegraphics[width=5cm]{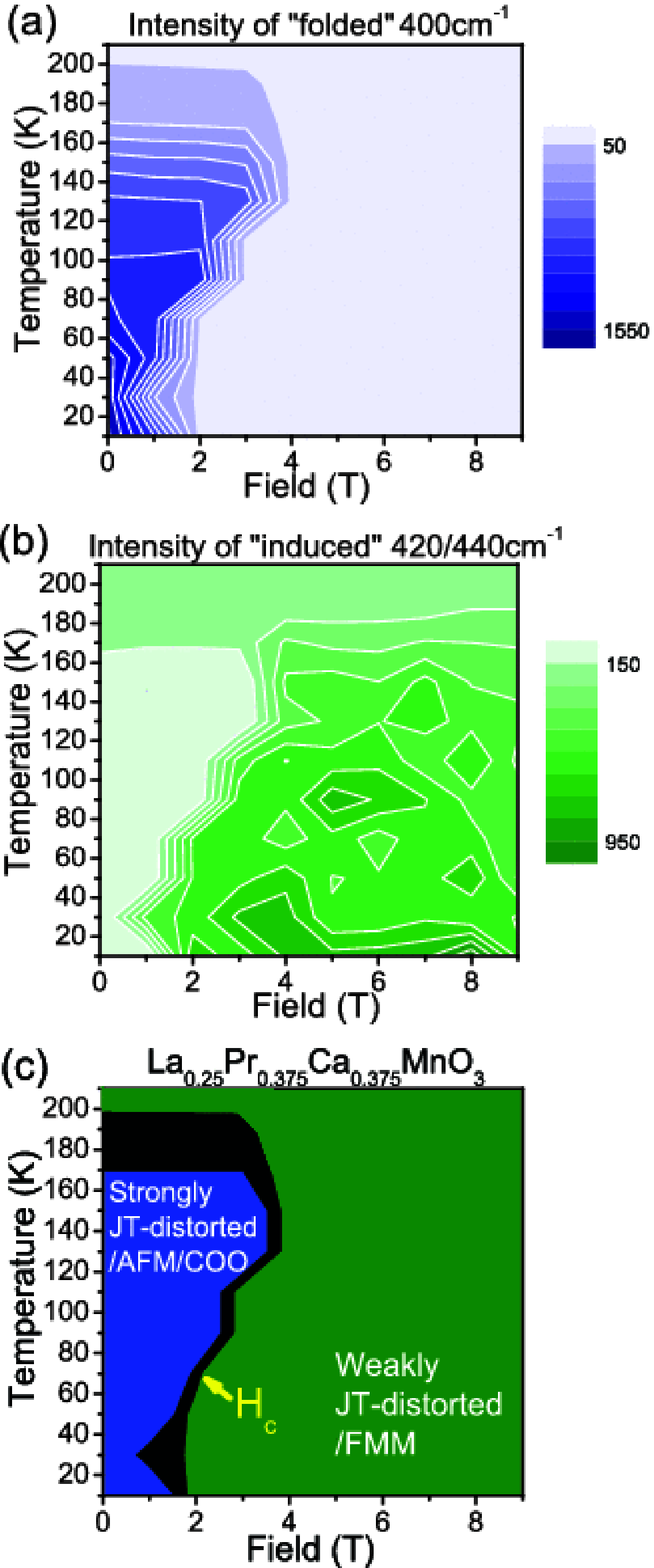}}
 \caption{\label{fig8}(a) Contour plot of the intensity of the 400 {\icm} folded phonon mode
 in {\lpcmo} as functions of temperature and field.
 The dark (purple) shades indicate the prevalence of the strongly JT distorted ($P2_1/m$ space group) structural phase,
 which is associated with the AFM/COO phase.
 (b) Contour plot of the combined intensities of the 420 and 440 {\icm} field-induced modes
 of {\lpcmo} as functions of temperature and field.  The dark (green) shades indicate the prevalence of the weakly JT
distorted
 ($Imma$ or $Pnma$ space group) structural phase, which is associated with the ferromagnetic metal (FMM) phase.
 (c) Temperature/magnetic field phase diagram of {\lpcmo}, inferred from the intensity plots
 in (a) and (b), illustrating strongly JT distorted ($P2_1/m$ space group) AFM/COO phase (purple),
 weakly JT distorted ($Imma$ or $Pnma$ space group) FMM phase (green), and coexistent crossover phase (black) regimes.}
\end{figure}

\subsubsection{Discussion}
Field-induced and thermal melting of COO in {\lpcmo} exhibit two
noteworthy differences compared with melting of COO in {\lcmo}:  (i)
Pr-substitution causes a remarkable disparity between field-induced
and thermal melting of COO in {\lpcmo} that is not observed in
{\lcmo}; more specifically, thermal melting of COO in {\lpcmo} shows
little difference from that observed in {\lcmo}, but field-induced
melting of COO in {\lpcmo} is substantially more abrupt---and occurs
at significantly lower fields---than field-induced melting of COO in
{\lcmo}; and (ii) Pr-substitution causes a dramatic reduction in the
size of the coexistent phase regime (black regions in Figs. 5(c) and
8(c)) in the $HT$-structural phase diagram compared to that observed
in {\lcmo}.  This leads to a significantly more abrupt field-induced
transition in {\lpcmo} between the strongly JT distorted ($P2_1/m$
space group) AFM/COO phase (purple region in Fig. 8(c)) and weakly
JT distorted ($Imma$ or $Pnma$ space group) FMM phase (green region
in Fig. 8(c)).

To explain this behavior, we note first that earlier work on
{\lpcmo} by Uehara et al. reported the coexistence of long-range FM
domains and long-range COO domains for $T$$<$80 K and $H$=0 T, and
the coexistence of short-range COO domains and short-range FM
domains for 80$<$T$<$210 K and $H$=0 T.~\cite{uehara99} Accordingly,
the dramatic difference between thermal and field-induced melting we
observe in {\lpcmo} indicates that---by favoring FM domains at the
expense of COO domains---applied magnetic fields are far more
effective than thermal fluctuations at disrupting long-range
coherence between COO domains and driving a COO/AFM to FM
transition, in {\lpcmo}.  This description is consistent with the
observed increase with temperature in the critical field $H_c$($T$)
at which {\lpcmo} transitions from the COO/AFM regime to the weakly
JT distorted FM regime (see Fig. 8(c)): with increasing temperature,
the size of the FM domains in {\lpcmo} is known to decrease to
nanometer length scales,~\cite{uehara99} thereby creating more PM
regions and a higher critical field needed to achieve a percolation
transition of the FM domains in the FM ``weakly JT distorted''
phase.  Thus, we argue that field-induced melting of long-range COO
is fundamentally different in {\lpcmo} and {\lcmo}:  in {\lcmo},
field-induced melting of COO primarily proceeds by a gradual
reduction of COO domain regions.  By contrast, in {\lpcmo}, the more
rapid field-induced melting of COO occurs via the enhancement of FM
domain regions that are already present in the material---even at
low temperatures and $H$=0 T---by Pr-substitution. However, thermal
melting in {\lpcmo} and {\lcmo} proceeds in the same fashion, namely
by disrupting COO domain regions and increasing the volume of the PM
phase.

These results further suggest that Griffiths phase
models~\cite{griffths69,bray87,chan06} may be appropriate
descriptions of the disordered {\lpcmo} system.  The Griffiths phase
describes the behavior of a random magnetic system that is between
the completely disordered and ordered magnetic phases,~\cite{bray87}
and was first used to describe a randomly dilute Ising ferromagnet
in the regime between the observed (suppressed) FM critical
temperature $T_C$ and the critical temperature $T_G$ of the pure
magnet (i.e., at which temperature the magnetization is non-analytic
in an external field).~\cite{griffths69}  This suppression of the
critical temperature was associated in the Griffiths phase with the
effects of quenched disorder, which partition the pure system into
small ferromagnetic clusters.  There have been several recent
studies that identify a Griffiths phase in the
manganites.~\cite{chan06,deisenhofer05}  In particular, we note that
the $T$-$x$ phase diagram of La$_{1-x}$Sr$_x$MnO$_3$ reported by
Deisenhofer et al.,~\cite{deisenhofer05} in which a Griffiths phase
regime is identified, is quite similar to the $T$-$y$ phase diagram
of La$_{0.625-y}$Pr$_y$Ca$_{0.375}$MnO$_3$,~\cite{uehara99} in which
a coexistence regime of FM and CO phases having different length
scales is reported.  Indeed, based on this similarity, we can make a
rough correspondence between the phases of La$_{1-x}$Sr$_x$MnO$_3$
and La$_{0.625-y}$Pr$_y$Ca$_{0.375}$MnO$_3$.  The correspondence
between the phases observed in these materials is summarized in  Table~\ref{tab:table1}.

\begin{table*}
\caption{\label{tab:table1}This table summarizes the corresponding phases among the $T$-$x$ phase diagram of La$_{1-x}$Sr$_x$MnO$_3$, the $T$-$y$ phase diagram of La$_{0.625-y}$Pr$_y$Ca$_{0.375}$MnO$_3$, and the $T$-$H$ phase diagram of {\lpcmo}.}
\begin{ruledtabular}
\begin{tabular}{lclcc}
 \multicolumn{2}{c}{La$_{1-x}$Sr$_x$MnO$_3$~\footnote{Ref.~\onlinecite{deisenhofer05}}}
 &\multicolumn{2}{c}{La$_{0.625-y}$Pr$_y$Ca$_{0.375}$MnO$_3$~\footnote{Ref.~\onlinecite{uehara99}}}
 &{\lpcmo}~\footnote{From our discussion}\\ \hline

 Griffiths phase&{0.075$\leq$$x$$\leq$0.16}&short-range CO/FM phase&{0.275$\leq$$y$$\leq$0.40}
&{$\sim$80 K$\leq$$T$$\leq$$\sim$200 K}\\
 &{$T_C(x)$$<$$T$$<$$T_G$}&&{$T_C(y)$$<$$T$$<$$T_{CO}$}
&{0T$<$$H$$<$$H_c(T)$}\\ \hline

 FM/insulating phase&{0.075$\leq$$x$$\leq$0.16}&long-range CO/FM phase&{0.275$\leq$$y$$\leq$0.40}
&{0K$\leq$$T$$\leq$$\sim$80K}\\
 &{0K$<$$T$$<$$T_C(x)$}&&{$T_C(y)$$<$$T$$<$$T_{CO}$}
&{0T$<$$H$$<$$H_c(T)$}\\ \hline

 FM/metallic phase&{0.16$<$$x$}&long-range FM&{$y$$<$0.275}&{$H_c(T)$$<$$H$}\\
 &{$T$$<$$T_C(x)$}&/short-range CO phase&{$T$$<$$T_C(y)$}\\

\end{tabular}
\end{ruledtabular}
\end{table*}

Therefore, we propose that the field- and temperature-dependent
behavior observed in {\lpcmo} can be understood in the context of
Griffiths physics:  In this picture, $T_C$ is dependent on the size
of the ferromagnetic clusters introduced by disorder, i.e., the
cluster size scales with $T_C$.  This is consistent with our
observation that the transition from COO to FM in {\lpcmo} can be
readily induced by applying even relatively weak fields at low
temperatures, where larger FM regions coexist with COO regions.

Further, the abrupt field-induced transition between COO and FM
phases in {\lpcmo}---which is in contrast to the more gradual
transition observed in {\lcmo}---is naturally explained in this
description as due to a percolation transition of the FM clusters,
which can be associated with a non-analyticity of the magnetization
in the Griffiths model.~\cite{chan06}  However, we hasten to add
that there has been no systematic theoretical study of the
field-dependence of ``Griffiths phase'' systems, and more
theoretical study would be welcome to check the connection between
the novel field-induced melting behavior observed in {\lpcmo} and
that expected from Griffiths phase physics.

\section{Summary and Conclusions}
Our detailed temperature- and magnetic-field-dependent Raman study
of the phase transitions in {\lcmo} and {\lpcmo} allows us to
provide specific details regarding the various structural phases
that accompany thermal and field-induced melting of COO in these
materials.  In {\lcmo}, we have found that thermal melting exhibits
three distinct temperature regimes:  below 150 K, melting of COO
occurs at the interface between large coherent COO regions ($P2_1/m$
space group), causing a reduction in the COO domain volume with
increasing temperature, but maintaining the long-range coherence of
COO domains.  Between 150$\leq$$T$$\leq$170 K, additional melting
leads to the eventual collapse of long-range COO, and the evolution
of a coexistence regime consisting of short-range COO ($P2_1/m$
space group) domains and ``weakly JT distorted'' ($Imma$ or $Pnma$
space group) regions.  Above 170 K, COO domains are completely
melted, leading to a weakly JT distorted ($Imma$ or $Pnma$ space
group) regime.  This temperature-dependent evolution of the spectra
is suggestive of a first-order ``electronic crystallization''
transition in {\lcmo}, with the kinetics in the coexistence region
corresponding to the propagation of a phase front.  The
field-induced melting process of COO in {\lcmo} is found to be quite
similar to the thermal melting process, suggesting that
field-induced melting of COO in {\lcmo} is associated with a
first-order quantum phase transition:  the application of a magnetic
field disrupts antiferromagnetic order preferentially at the surface
of COO regions, giving rise to heterogeneous melting of the coherent
long-range COO regions, and a three-stage field-dependent melting
process:  a low-field regime in which a long-range ``strongly JT
distorted''/COO phase is present; an intermediate-field regime in
which strongly JT distorted/COO domain regions coexist with weakly
JT distorted/FMM domain regions; and finally a high-field regime in
which a long-range ``weakly JT distorted''/FMM phase associated with
either a $Imma$ or $Pnma$ structure is present.  This three-stage
field-dependent melting process in {\lcmo} is consistent with the
three distinct structural phases identified by Tyson et
al.~\cite{tyson04} over a more limited temperature range.  Based on
our results, we were able to obtain a complete phase diagram of
{\lcmo} in the temperature range 6--170 K and in the field range
0--9 T.  Notably, the similarity between thermal and field-induced
melting in {\lcmo} is reflected in critical field values,
$H^*(T)$---between the long-range COO and intermediate
``coexistence'' regimes---and $H_c(T)$---between the intermediate
``coexistence'' and weakly JT distorted FMM regimes---that decrease
roughly linearly with decreasing temperature at a rate of
$\sim$$-$0.2 T/K.  Further, our results suggest an approximate $T$=0
critical point of $H^*(T$=$0)$$\sim$30 T for {\lcmo}, indicative of
a very robust COO in {\lcmo}.

To investigate the effects of disorder on commensurate COO in
{\lcmo}, we also examined the {\lpcmo} system.  We found that while
thermal melting in {\lpcmo} is not significantly different from that
in {\lcmo}, field-induced melting in {\lpcmo} differs dramatically
from that in {\lcmo}:  the application of a magnetic field in the
former material is found to induce an abrupt transition from a
long-range COO to a ``weakly JT distorted''/FM phase with a very
narrow intermediate coexistence field regime, in contrast to a
gradual transition observed in {\lcmo}.  Moreover, the phase diagram
deduced for {\lpcmo} reveals that the critical field $H_c(T)$
between COO and weakly JT distorted regimes increases from $\sim$2 T
at $T$=3.5 K to 4 T at $T$=130 K, in contrast to a linear decrease
of the $H_c(T)$ with decreasing temperature in {\lcmo}.  This
suggests that disordering COO with Pr-substitution does not
appreciably affect temperature-dependent melting of COO, but does
significantly affect the field-induced melting of COO.  We argue
that the magnetic field induced coherence between FM domains---which
originate from the introduction of Pr in {\lpcmo}---accelerates the
disruption of antiferromagnetic order in COO, and leads to a
distinctly different process of field-induced melting of COO from
that observed in {\lcmo}.  In particular, we argue that
field-induced melting of COO in {\lcmo} is best described as the
percolation of FM domains introduced at $H$=0 T by Pr-substitution,
and we suggest the Griffiths phase physics may be an appropriate
theoretical model for describing the unusual temperature- and
field-dependent transitions observed in {\lpcmo}.

\begin{acknowledgments}
This material is based on work supported by the U.S. Department of
Energy, Division of Materials Sciences, under Award No.
DE-FG02-07ER46453, through the Frederick Seitz Materials Research
Laboratory at the University of Illinois at Urbana-Champaign.  Work
at Hamburg was supported by the German funding agency under RU
773/3-1, and work at Rutgers was supported by NSF-DMR-0405682.
\end{acknowledgments}

\end{document}